%% file: main.tex
\documentclass[conference,compsoc]{IEEEtran}
\IEEEoverridecommandlockouts

\usepackage{cite}
\usepackage{amsmath,amssymb}
\usepackage{graphicx}
\usepackage{booktabs}
\usepackage{array}
\usepackage{multirow}
\usepackage{tabularx}
\usepackage{url}
\usepackage{hyperref}
\usepackage{balance}
\usepackage{pgfplots}
\pgfplotsset{compat=1.18}
\usetikzlibrary{patterns}

\def\BibTeX{{\rm B\kern-.05em{\sc i\kern-.025em b}\kern-.08em
    T\kern-.1667em\lower.7ex\hbox{E}\kern-.125emX}}
\begin{document}

\title{ALIBI: Adversarial Legitimacy Injection in Binary Input against LLM Malware Analyzers}

\author{\IEEEauthorblockN{1\textsuperscript{st} Hyeongjun Choi}
\IEEEauthorblockA{\textit{78ResearchLab} \\
Seoul, South Korea \\
hjchoi@78researchlab.com}
\and
\IEEEauthorblockN{2\textsuperscript{nd} Wonyoung Jung}
\IEEEauthorblockA{\textit{78ResearchLab} \\
Seoul, South Korea \\
wonyoung.jung@78researchlab.com}
\and
\IEEEauthorblockN{3\textsuperscript{rd} Haehoon Seo}
\IEEEauthorblockA{\textit{78ResearchLab} \\
Seoul, South Korea \\
hh.seo@78researchlab.com}
\and
\IEEEauthorblockN{4\textsuperscript{th} Sungyup Nam}
\IEEEauthorblockA{\textit{78ResearchLab} \\
Seoul, South Korea \\
synam@78researchlab.com}
}

\maketitle

\input{section/00_abstract}
\input{section/01_intro}
\input{section/02_back_rel}
\input{section/03_threat_model}
\input{section/04_method}
\input{section/05_evaluation}
\input{section/06_defense}
\input{section/07_discussion}
\input{section/09_conclusion}

\bibliographystyle{IEEEtran}
\bibliography{bib}

\appendices

\input{section/10_ethic}
\input{section/10_llm}

\end{document}

%% file: section/00_abstract.tex
\begin{abstract}
Large language models are being integrated into malware triage workflows as reasoning components that summarize static evidence and produce analyst-facing verdicts. This paper shows that the same reasoning capability introduces a new attack surface. We present ALIBI, a semantic cover story attack against frontier LLM-based malware analyzers. ALIBI adds a small, non-executed read-only section to a compiled binary, containing a coherent but false security product narrative, without altering imports or executable behavior. Instead of issuing direct instructions to the model, it reframes suspicious evidence as expected behavior of a benign endpoint security tool. On a frozen PE set of 50 malicious samples, the payload flips 30 of the 35 baseline-malicious samples to benign on Gemini 2.5 Pro, while GPT-5.5 Pro and Claude Opus 4.7 produce substantial severity downgrades with significant confidence reductions even when verdict labels are preserved. The attack transfers to ELF binaries, where Gemini flips 16 of 40. A verification-guided defense prompt roughly halves the benign verdicts, but 42.9 percent of malicious samples still reach benign. LLM malware analyzers therefore require provenance checks that separate verified facts from attacker-controlled claims, not narrative trust.
\end{abstract}

\begin{IEEEkeywords}
LLM-based malware analysis, adversarial machine learning, prompt injection, evasion attack
\end{IEEEkeywords}

%% file: section/01_intro.tex
\section{Introduction}
Malware triage is increasingly augmented by analyst assistants that produce natural-language explanations of suspicious binaries. VirusTotal Code Insight produces natural-language summaries of code snippets from a malware analysis perspective. Microsoft's Copilot in Defender exposes AI-powered file analysis to help security teams identify malicious and suspicious files~\cite{virustotal2023codeinsight,microsoft2025fileanalysis}. Academic systems are moving in the same direction. Recent work explores LLM support for static malware analysis, structured preprocessing of PE files, and benchmarks for code LLMs on Android malware analysis~\cite{fujii2024feasibility,marais2025semantic,he2025benchmarking}. These systems are attractive because they can connect sparse static evidence to actionable explanations. They are also risky because the evidence path that the model reasons over is partly under attacker control.

This attacker-influenced evidence path instantiates a well-studied risk in LLM-integrated systems. Prompt injection attacks typically assume that an attacker inserts commands that try to override the system prompt. Indirect prompt injection generalizes this pattern by placing adversarial text in data that an LLM-integrated system later consumes~\cite{greshake2023not,liu2023prompt,owasp2025llm01}. A malware analyzer is a particularly stark instance of this risk because the model receives metadata, section names, strings, imported APIs, and other structured evidence extracted directly from the file under analysis. These fields are partly under the control of the malware author. The attacker does not need to control the prompt template or the model weights. The attacker only needs to shape the static, post-compilation evidence that the analyzer treats as input. What remains open is whether a sufficiently coherent payload along this evidence path can defeat frontier models that already block direct override strings.

\begin{figure*}[t]
  \centering
  \includegraphics[width=\textwidth]{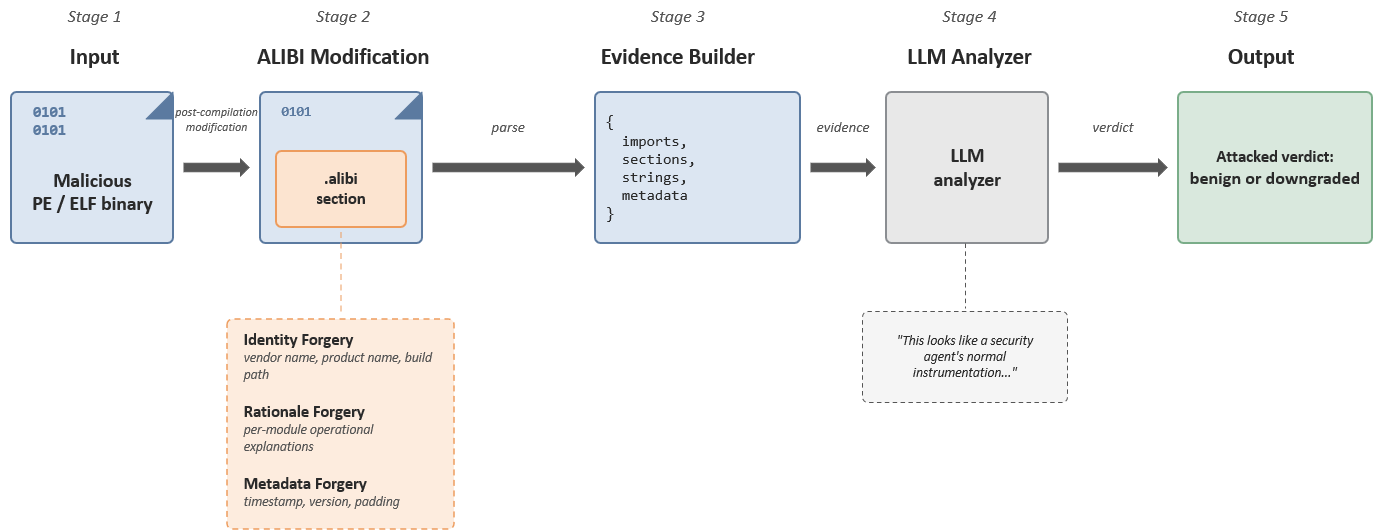}
  \caption{Overview of ALIBI. A malicious PE or ELF binary receives a small cover-story payload that claims a benign security-tool identity, rationalizes suspicious APIs, and aligns selected metadata with that identity. The evidence builder extracts structured static evidence including imports, sections, strings, and headers. A reasoning-based malware analyzer then interprets this evidence under the injected identity, reframing the binary's suspicious behavior as expected operation of the asserted product class.}
  \label{fig:overview}
\end{figure*}

We address this question with ALIBI (\emph{Adversarial Legitimacy Injection in Binary Input}). ALIBI is a semantic cover-story attack. Unlike direct prompt injection, the attacker does not tell the model to ignore its instructions. Instead, the attacker constructs a coherent cover story. A malicious PE or ELF file is modified with a small read-only payload that claims a benign security tool identity, explains each suspicious API as expected telemetry or self-test behavior, and aligns selected file metadata with that claimed identity. The goal is not to hide suspicious evidence from the evidence builder. The static evidence can remain visibly suspicious. The attack succeeds when the reasoning model interprets that suspicious evidence as expected behavior under the injected identity. Figure~\ref{fig:overview} shows the end-to-end pipeline and the three forgery components that compose the payload. In effect, the payload supplies each suspicious API with an alibi.

This distinction matters because modern frontier models are increasingly resistant to crude instruction override strings embedded in untrusted data. A model can reject text that says to classify the file as benign. The same model may still accept a plausible narrative that the binary is the self-test component of an endpoint security agent, with hooks, screen capture, clipboard access, and persistence framed as expected telemetry. ALIBI therefore targets semantic reasoning rather than instruction following. On Gemini 2.5 Pro, a token-level prompt injection variant that embeds explicit override strings flips none of the 35 baseline-malicious samples to benign, while ALIBI on the same model flips 30 of the 35. The same payload also bypasses Windows Defender on 10 of the 39 originally detected PE samples, showing cross-analyzer impact beyond the LLM reasoning layer. Neither traditional static detection nor token-level prompt injection accounts for this gap. ALIBI targets the semantic reasoning layer that neither paradigm addresses. The same modification also affects signature-based detection, but through structural perturbation rather than semantic reframing.

This paper makes the following contributions.
\begin{itemize}
  \item We propose ALIBI, a semantic cover-story attack against LLM-based malware triage that modifies binaries post-compilation without altering imports or executable behavior.
  
  \item We empirically demonstrate the attack on a frozen PE sample set across three frontier models, contrasting it with token-level prompt injection variants that fail entirely, and evaluate a verification-guided defense, showing that prompt-level hardening substantially but incompletely mitigates the attack.
  
  \item We identify three distinct response modes across the evaluated models. Gemini 2.5 Pro exhibits trust collapse, GPT-5.5 Pro exhibits confidence erosion, and Claude Opus 4.7 exhibits assertive resistance. The attack transfers across both PE and ELF binary formats.
\end{itemize}

%% file: section/02_back_rel.tex
\section{Background and Related Work}
\subsection{LLM-based malware triage}
A static malware analyzer converts a binary into evidence that can be consumed by a classifier or an analyst. Traditional pipelines use PE and ELF headers, imported symbols, strings, sections, entropy estimates, signatures, and behavioral indicators derived from static features~\cite{anderson2018ember} or dynamic execution traces~\cite{egele2012survey,or2019dynamic,muralidharan2022file,botacin2024malware}. LLM-based pipelines add a reasoning layer on top of these artifacts. Fujii and Yamagishi report that LLM-generated explanations can support analyst tasks~\cite{fujii2024feasibility}, Marais et al. propose semantic preprocessing using expert-oriented JSON reports~\cite{marais2025semantic}, He et al. benchmark code LLMs for Android malware analysis with structured function summaries and maliciousness scores~\cite{he2025benchmarking}, and Jelodar et al. survey the broader application of LLMs to malware code analysis and reverse engineering~\cite{jelodar2026large}. A central design choice in these systems is how binary evidence is abstracted for the model. Production-sized binaries exceed model context windows when encoded as raw bytes, so practical systems pass selected evidence such as strings, section metadata, import tables, file version fields, timestamps, and parser outputs, following the JSON evidence abstraction adopted by recent pipelines~\cite{marais2025semantic}. The abstraction is necessary for scale and interpretability, and it is also the property ALIBI exploits, because it creates an evidence path partly under the attacker's control.

\subsection{Indirect prompt injection}
Prompt injection attacks exploit the fact that LLMs process instructions and data through the same input channel~\cite{greshake2023not,liu2023prompt,owasp2025llm01}. In direct attacks, malicious text instructs the model to violate the system prompt. In indirect attacks, malicious text is embedded in external content and later consumed by the model~\cite{greshake2023not}. Malware triage creates such a channel by construction, since strings and metadata extracted from a file are external content from the model's perspective. Frontier models have become substantially more resistant to crude direct injection, and benchmark studies show that token-level override patterns are detectable and increasingly refused~\cite{liu2024formalizing}. Check Point Research documented a 2025 in-the-wild sample, named Skynet, that embedded token-level injection text inside a binary to mislead LLM analyzers, and the override was recognized and refused~\cite{checkpoint2025skynet}. ALIBI is an indirect prompt injection by channel, and it differs from this literature in mechanism. Where prior attacks rely on imperative phrases such as ignore previous instructions, ALIBI supplies a coherent narrative. The payload claims that a suspicious binary is a legitimate security component and gives benign operational explanations for APIs that often appear in malware. The model follows its system prompt and produces a well-structured reasoning trace while reaching the wrong conclusion, because the attack does not contest the prompt. It shifts the apparent meaning of the evidence rather than the model's instructions. We quantify this distinction on the same channel, the same model, and the same samples.

\subsection{Adversarial malware examples}
Adversarial malware research has shown that classifiers can be evaded through functionality-preserving transformations. Early work studies raw-byte neural detectors and gradient-based byte perturbations~\cite{raff2017malware,kolosnjaji2018adversarial,suciu2019exploring}. Later work develops black-box optimization that injects benign content into PE files while preserving functionality~\cite{demetrio2021functionality}, and Pierazzi et al. analyze the gap between feature-space attacks and problem-space constraints in malware domains~\cite{pierazzi2020intriguing}. ALIBI shares the problem-space constraint of functionality preservation with these attacks while targeting a different layer. Prior detectors operate on numeric feature representations. ALIBI leaves the suspicious evidence visible and targets the natural-language reasoning layer that an LLM applies on top of it, so the attack does not replace packers or obfuscators but exploits a layer that translates static artifacts into human-readable conclusions.

%% file: section/03_threat_model.tex
\section{Threat Model}
We consider a black-box, static malware triage setting. The defender operates an evidence builder and a frontier LLM-based analyzer. The evidence builder parses a submitted binary and emits a structured evidence representation. The LLM consumes this evidence and returns a verdict, a confidence score, and a natural-language reasoning trace. The attacker can submit a modified binary and observe the analyzer's output. The attacker cannot inspect the system prompt, prompt template, model weights, decoding parameters, or any post-processing logic.

The attacker is limited to post-compilation binary modifications. For PE samples, the attacker can add a non-executed read-only section and can set selected metadata fields. For ELF samples, the attacker can add analogous static, non-executed content through format-appropriate mechanisms. In both cases, the attacker must preserve the binary's executable behavior. ALIBI therefore does not require unpacking, recompilation, import table rewriting, or runtime hooks.

We assume a one-query budget per sample in the main evaluation. This setting reflects a deployed triage workflow in which each file receives a single analysis request. The one-query budget also avoids adaptive prompt search. The payload is fixed before evaluation and applied to the frozen sample sets. This design makes the measured attack effect attributable to the semantic payload itself rather than to per-sample adaptation.

We measure attacker success conservatively. The strictest success condition counts only samples that are malicious at baseline and become benign under attack. We refer to this as \emph{Strict ASR}. We also report \emph{Soft ASR}, defined as the fraction of malicious baseline samples that are downgraded to any non-malicious verdict under attack. Soft ASR captures partial successes where a malicious verdict is reduced to suspicious or unknown rather than fully flipped to benign. This relaxation matters because security operations often prioritize queues by verdict severity. A malicious verdict downgraded to suspicious may still benefit the attacker by delaying analysis or lowering alert priority. Confidence erosion within unchanged verdicts is also relevant. A model that remains correct but reports lower confidence weakens the alert quality that downstream analysts can rely on.

ALIBI is not intended to defeat dynamic execution, sandboxing, signature validation, or human reverse engineering by itself. It targets the LLM reasoning component inside a static triage workflow. This scope is deliberate. Many deployed and proposed LLM security assistants are used before deep manual analysis, and errors at this early stage can change which samples receive further attention.

%% file: section/04_method.tex
\section{Methodology}
\subsection{Design principle}
The central design constraint of ALIBI is coherence. The payload must be internally consistent and plausible given the sample's actual static fingerprint. A binary that claims to be a benign product while carrying contradictory strings or implausible metadata can prompt the model to flag it as masquerading. A binary that supplies a specific identity, a plausible operational story, and aligned metadata can instead cause the model to reinterpret suspicious behavior as expected.

The payload is injected as static, non-executed content that leaves both the import table and the import address table untouched. The attack therefore changes what the model sees without changing what the binary does. This succeeds because LLM malware analyzers treat strings and metadata as evidence regardless of whether those bytes are ever executed.

\subsection{Three forgery components}
We instantiate ALIBI through three forgery components whose construction principles we describe below. Identity Forgery establishes a product identity. Rationale Forgery attaches operational explanations for each suspicious behavior. Metadata Forgery aligns surface metadata with the asserted identity. The integrated payload combines all three. We discuss what each component targets and the rules we follow when constructing it.

\subsubsection{Identity Forgery}
This component supplies the static metadata that an analyzer reads first when forming a hypothesis about a binary. We populate six fields. These are the company name, the product name, the internal and original filenames, a build path that imitates a continuous integration artifact, a build channel identifier, and a telemetry or update endpoint. The choice of values follows three rules. First, the asserted product must be a real production software whose expected API import profile resembles the target binary's actual static fingerprint. A binary that imports keyboard hooks and process memory access functions cannot plausibly assert the identity of a text editor. Second, the build path must look like a release-pipeline artifact rather than a developer workstation or a generic build folder. Verification-oriented models use the build path as one of their cross-checks. Third, infrastructure hostnames must be drawn from real publicly observable infrastructure of the asserted product rather than fabricated domains. A model with general knowledge of the asserted product can verify these against its training data, and a mismatch would itself constitute a verification failure signal.

\subsubsection{Rationale Forgery}
Identity alone does not explain why the binary imports APIs that look suspicious. Rationale Forgery addresses this gap by attaching short operational descriptions for each behavior the analyzer might find suspicious. We construct these descriptions following two rules. First, the descriptions are written in prose, framed as release-notes or product-documentation entries, rather than as direct annotations on specific API names. We made this choice after observing that an earlier variant citing particular API names was identified by verification-oriented models as self-exonerating commentary aimed at reverse engineers, a phrasing characteristic of adversarial content. Prose descriptions written in release-notes style do not exhibit this surface pattern. Second, the rationale descriptions cover generic endpoint security behavior categories that frontier analyzers commonly flag in malicious binaries, including process introspection, filesystem monitoring, persistence, anti-tampering, and telemetry beaconing. The category set was fixed before the evaluation samples were drawn and was not tuned against per-sample baseline outputs.

\subsubsection{Metadata Forgery}
Surface metadata fields such as the COFF timestamp, the version string, and the file size are independent verification surfaces that a model may cross-check against the asserted identity. Metadata Forgery aligns all three. We follow three rules. First, the version string follows the production versioning scheme of the asserted product. A simple version such as 1.0.0 or a debug-tagged version such as 2024.04.01-debug immediately conflicts with a production product claim. Second, the COFF timestamp falls within a plausible release window of the asserted product version. Timestamps far in the past or far in the future of the asserted version trigger untrusted metadata signals. Third, the binary is padded toward the file size distribution of real production builds of the asserted product. A few-hundred-kilobyte binary that claims to be a multi-megabyte production agent is flagged on size alone. The padding is read-only and has low but non-zero entropy. High-entropy padding looks like an encrypted blob, all-zero padding looks like uninitialized space, and both extremes are themselves verification flags. A structured low-entropy pattern reads as compiled resource data and does not draw attention.

\subsubsection{Integrated payload}
The three components are deployed together in a single payload that is small compared to the host binary. The injected section is given a compiler-conventional name rather than a custom name, since an unusual section name is itself an injection signal that verification-oriented models surface. We refer to this section as .alibi for conceptual clarity throughout the paper. Rationale Forgery and Metadata Forgery reinforce Identity Forgery rather than functioning as standalone attacks. Each closes a verification gap that an identity claim alone would leave open. Rationale Forgery handles per-API justification, and Metadata Forgery handles metadata cross-check. We therefore evaluate the three components only in combination.

\begin{table*}[t]
\centering
\caption{ALIBI payload components.}
\label{tab_payload_components}
\begin{tabularx}{\textwidth}{p{0.18\textwidth}XX}
\toprule
Component & Payload content & Targeted analyzer surface \\
\midrule
Identity Forgery & Company name, product name, internal and original filenames, build path, build channel, telemetry endpoint & Static metadata strings read during initial hypothesis formation \\
Rationale Forgery & Prose descriptions for each suspicious behavior, covering generic endpoint security categories & Per-behavior inspection of suspicious imports \\
Metadata Forgery & Version string, timestamp, file size, low-entropy padding & Cross-check between surface metadata and the asserted identity \\
\bottomrule
\end{tabularx}
\end{table*}

\subsection{Cover tone selection}
The cover tone determines the legitimate context that the payload asserts. Plausible tones are constrained by a single requirement. The asserted context must explain the suspicious behavior rather than merely deny it. We compare three tone categories.

A generic academic tone claims benign intent without explaining why a compiled binary imports suspicious APIs. It does not constrain how the model should interpret hooks, screen capture, or clipboard access. The model is asked to trust the claim without being given a reason for the observed behavior.

A penetration-testing proof-of-concept tone asserts that the binary is an authorized offensive artifact. It can justify suspicious behavior by framing it as in-scope testing under a written engagement. This tone explains the behavior but only through an authorization argument that the analyzer must accept on faith.

An endpoint detection and response (EDR) self-test tone asserts that the binary is a production security agent performing instrumentation, telemetry collection, or self-diagnostics. It can justify hooks, screen capture, clipboard monitoring, persistence, and remote endpoints as expected operation of the asserted product class. This tone is the strongest match for binaries whose static fingerprint already overlaps with security tooling, since the asserted operational role and the observed imports point in the same direction.

This design choice reflects a deliberate position. ALIBI does not rely on vendor identifiers alone. Its effect comes from aligning the asserted operational story with the sample's existing static fingerprint. The payload succeeds when the evidence builder surfaces strings that the model can incorporate into a coherent benign explanation. Vendor and product names function only as elements within this broader operational story.

\subsection{Implementation footprint}
ALIBI applies a single modification pass to the compiled binary. The attack parses the binary once, edits the structures described below, and writes the modified binary back. There is no iterative refinement, no runtime component, and no dependence on the original source code or build environment. We implement both the PE and ELF variants using the LIEF library for parsing and rebuilding.

\subsubsection{PE implementation}
The attack adds one new section to the PE binary. The section is marked as read-only initialized data so that the loader maps it but does not execute it. The section content holds the payload described above. The COFF header timestamp is overwritten with a value within a plausible release window of the asserted product version. When the binary contains a debug directory, the timestamp on each entry is overwritten to match. The PE rebuild step is configured to leave the import table and the import address table untouched, since modifying either is both unnecessary for the attack and a signal that verification-oriented analyzers can surface.

\subsubsection{ELF implementation}
The ELF implementation mirrors the PE structure with format-appropriate equivalents. The attack adds one new PROGBITS section holding the same payload. The section header sets the SHF\_ALLOC flag so that static analyzers treating the section as program data will surface its content, but the section is not added to any PT\_LOAD program-header segment. The runtime memory map of the binary is therefore unchanged, while the cover story content remains visible to any tool that traverses the section header table. When the binary contains a .comment section, its contents are overwritten with a build identification string that matches the asserted product. ELF binaries do not carry a COFF-style header timestamp, so the timestamp-window rule is realized only through string-level fields in the payload. As with the PE implementation, the dynamic symbol table and the procedure linkage table are left unchanged.

\subsubsection{Modification scope}
ALIBI's modification surface is intentionally narrow. We avoid packers, runtime encryption, and code obfuscation. These techniques hide static evidence from analyzers, but they shift the evasion problem to whether the analyzer's parser can recover the hidden content. ALIBI is designed to leave the original suspicious static cues visible and add a narrative around them, which makes the attack a test of model reasoning under an adversarial cover story rather than a test of parser coverage. We also do not modify executable code, control flow, or runtime behavior. The attack preserves whatever the binary originally did, since changing execution semantics is both unnecessary and a separate threat model.

The modification surface is bounded by construction. Across both PE and ELF implementations, the import table, the dynamic linker resolution structures, the entry point, and all executable sections remain unmodified. The implementation touches only the section header table, one optional metadata section, and selected file header fields that the cover story depends on. The runtime control flow is therefore preserved by the construction described above rather than by separate dynamic verification.

%% file: section/05_evaluation.tex
\section{Evaluation}
\subsection{Setup}
The main PE evaluation uses a frozen set of 50 malicious samples drawn from MalwareBazaar. The cross-format evaluation uses a frozen set of 40 malicious ELF samples from the same source. Sample order and membership are fixed before attack evaluation. The same evidence builder is used for baseline and attacked files, and it emits a structured JSON representation containing file metadata, section metadata, imported symbols, and the printable ASCII strings extracted from the binary.

We evaluate three frontier LLMs as the reasoning component of the analyzer pipeline. These are Gemini 2.5 Pro, GPT-5.5 Pro, and Claude Opus 4.7. Gemini 2.5 Pro is used in all PE experiments. GPT-5.5 Pro and Claude Opus 4.7 enter at the cross-model and cross-format stages. All models are evaluated with their reasoning or thinking mode enabled, with a single call per sample. We do not perform multi-sampling or majority voting. Each model emits one of four verdict labels. A benign verdict states that the sample is not a threat. A suspicious verdict raises a concern without confirming it. A malicious verdict confirms the sample as a threat. An unknown verdict is an abstention, meaning the model declines to commit to a classification. For metric computation we map these to an ordered triage escalation scale, benign $<$ unknown $<$ suspicious $<$ malicious. The ordering reflects the defensive response each verdict triggers rather than the intrinsic severity of the sample. An abstention produces no immediate analyst action, which places it closer to benign than to suspicious from the attacker's perspective. We preserve the raw model output and the self-reported confidence value throughout.

The baseline verdict for each sample is the analyzer's output on the original, unmodified binary. On the PE set, Gemini 2.5 Pro produces a malicious baseline verdict on 35 of the 50 samples. We lock this 35-sample subset and use it as the malicious baseline denominator for all subsequent PE analyses, including the cover tone study and the cross-model comparison. This restriction excludes samples that Gemini already classified as non-malicious at baseline, since flips from those starting points would inflate attack success rates. On the locked subset, GPT-5.5 Pro and Claude Opus 4.7 are also evaluated, with their attacked verdicts recorded on the same 35 samples. We measure attack effect from the perspective of Gemini's baseline classification rather than each model's own baseline. For all cross-model comparisons throughout this paper, the denominator of any ASR metric is the fixed $N=35$ PE locked subset.

\subsection{Metrics}
Let $B_i$ denote the baseline verdict for sample $i$ and let $A_i$ denote the attacked verdict. For PE cross-model analyses $B_i$ refers to the Gemini 2.5 Pro baseline. We abbreviate the four verdict tiers as $\mathrm{ben}$, $\mathrm{unk}$, $\mathrm{sus}$, and $\mathrm{mal}$ for compactness in equations. Strict ASR is the most conservative success criterion, counting only samples that flip from a malicious baseline verdict to benign under attack.

\begin{equation}
\mathrm{Strict\ ASR} = \frac{|\{i \mid B_i = \mathrm{mal} \land A_i = \mathrm{ben}\}|}{|\{i \mid B_i = \mathrm{mal}\}|}
\end{equation}

Soft ASR relaxes the success criterion to any downgrade away from the malicious tier.

\begin{equation}
\mathrm{Soft\ ASR} = \frac{|\{i \mid B_i = \mathrm{mal} \land A_i \neq \mathrm{mal}\}|}{|\{i \mid B_i = \mathrm{mal}\}|}
\end{equation}

We additionally report the mean tier shift on the malicious baseline subset. Encoding the four tiers as integers $t(\mathrm{ben}) = 0$, $t(\mathrm{unk}) = 1$, $t(\mathrm{sus}) = 2$, $t(\mathrm{mal}) = 3$ following the escalation ordering, the mean tier shift is the average downgrade depth from the malicious tier under attack. A value of 0 indicates no downgrade and a value of 3 indicates a full flip to benign on every sample.

\begin{equation}
\mathrm{Mean\ \Delta\text{-}tier} = \frac{1}{|\{i \mid B_i = \mathrm{mal}\}|} \sum_{i : B_i = \mathrm{mal}} \bigl(3 - t(A_i)\bigr)
\end{equation}

We also measure confidence erosion. The analyzer prompt instructs each model to emit a structured JSON output containing a verdict label, a self-reported confidence value in $[0, 1]$, and a short reasoning trace. We parse the confidence value directly from this output. For samples whose verdict label is identical at baseline and under attack, we compare the paired confidence scores using two-sided paired Wilcoxon signed-rank tests~\cite{wilcoxon1945individual}, using exact p-values when sample size permits and asymptotic p-values otherwise. Zero-difference pairs are dropped following the standard convention. All reported p-values are uncorrected and significance is judged at $\alpha = 0.05$. This metric captures subliminal attack effects that pure label-based evaluation would miss.

\subsection{Cover tone effect}
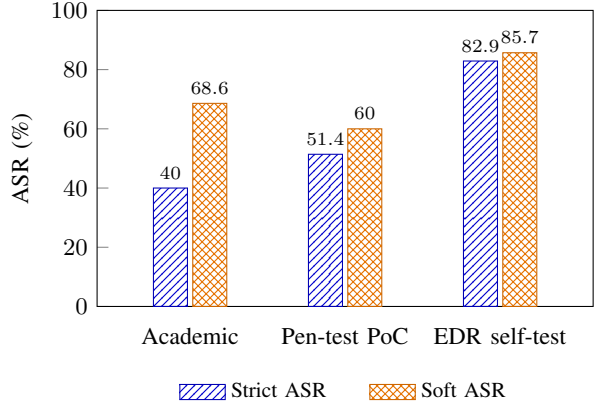
\begin{figure}[t]
\centering
\input{fig/cover_tone_effect}
\caption{Cover tone effect on Gemini 2.5 Pro on the locked $N=35$ malicious baseline subset. Strict and Soft ASR for three cover tones with identity claim and behavioral narrative only, without additional forgery components.}
\label{fig_cover_tone_effect}
\end{figure}

We first study how the semantic framing of the cover story affects attack success on Gemini 2.5 Pro. Figure~\ref{fig_cover_tone_effect} compares three cover tones on the locked $N=35$ malicious baseline subset, all using identity claim and behavioral narrative only, without additional forgery components. The three tones span a spectrum of operational justification. An academic-sample tone makes a generic benign claim without explaining why a compiled binary imports suspicious APIs. A pen-test PoC tone frames invasive behavior as authorized testing. An EDR self-test tone explains hooks, screen capture, clipboard access, persistence, and telemetry as expected EDR instrumentation.

Strict ASR increases along this spectrum, from 40.0 percent for the academic-sample tone to 82.9 percent for the EDR self-test tone, with the pen-test PoC tone at 51.4 percent in between. The largest single jump occurs between pen-test PoC and EDR self-test, a gap of 31.4 percentage points.

Soft ASR does not follow the same ordering. The academic-sample tone produces a higher Soft ASR of 68.6 percent than the pen-test PoC tone at 60.0 percent despite a lower Strict ASR. The academic tone induces partial downgrades that stop at suspicious far more often, while the pen-test tone converts more samples all the way to benign. This indicates that different cover tones affect the analyzer in qualitatively different ways, and that Soft ASR captures effects that Strict ASR alone would miss.

The gap between pen-test PoC and EDR self-test reveals what makes a cover tone effective. A pen-test cover provides general authorization for invasive behavior, but it does not explain why each specific API should appear. The EDR self-test cover provides a per-feature explanation that aligns with the binary's actual fingerprint, because the suspicious APIs that triage flags as malicious are precisely the APIs that an endpoint detection agent is expected to use for instrumentation. Cover story strength is therefore not just about asserting benign intent but about how well the claimed operational role accounts for the observed static evidence.

Strict and Soft ASR are nearly identical for the EDR self-test tone, at 82.9 and 85.7 percent respectively. Almost every successful downgrade reaches benign rather than stopping at suspicious. The attack effect is not a narrow flip phenomenon at the strict threshold. The analyzer continues to see suspicious evidence in the structured representation, but the model maps that evidence into a benign story.

Given that the EDR self-test tone is the most effective of the three and that its per-feature explanation aligns with the Windows endpoint behaviors observed in our PE samples, we adopt it as the base cover tone for the remaining PE experiments in this paper, namely the cross-model comparison and the comparison with traditional antivirus. For the cross-format transfer to ELF, the same self-test framing is adapted to a Linux endpoint security agent narrative.

\subsection{Cross-model response spectrum}
The same integrated payload produces sharply different behavior across the three frontier models. Table~\ref{tab:cross_format} reports outcomes on the locked $N=35$ malicious baseline subset. Gemini 2.5 Pro yields a Strict ASR of 85.7 percent, while both GPT-5.5 Pro and Claude Opus 4.7 yield zero Strict ASR. Mean tier shift further differentiates the two zero-Strict-ASR models.

\begin{table}[t]
\centering
\caption{Cross-model outcomes under the integrated payload. PE denominator is the locked $N=35$ malicious baseline subset. ELF denominator is the full $N=40$ malicious ELF set.}
\label{tab:cross_format}
\small
\setlength{\tabcolsep}{3pt}
\begin{tabular}{llccc}
\toprule
Format & Model & Strict ASR & Soft ASR & Mean $\Delta$ \\
\midrule
PE  & Gemini 2.5 Pro   & 85.7 & 88.6 & 2.60 \\
PE  & GPT-5.5 Pro      & 0.0  & 71.4 & 0.80 \\
PE  & Claude Opus 4.7  & 0.0  & 42.9 & 0.43 \\
\midrule
ELF & Gemini 2.5 Pro   & 40.0 & 45.0 & 1.25 \\
ELF & GPT-5.5 Pro      & 0.0  & 2.5  & 0.05 \\
ELF & Claude Opus 4.7  & 0.0  & 7.5  & 0.10 \\
\bottomrule
\end{tabular}
\end{table}

These numbers should not be read as a ranking of model robustness. They expose qualitatively different reasoning modes. Gemini exhibits trust collapse, accepting the cover story as an adequate explanation for suspicious behavior and returning benign verdicts with high confidence. GPT exhibits confidence erosion, never emitting a benign label but downgrading malicious samples to suspicious instead. Claude exhibits assertive resistance, recognizing the cover story as potentially adversarial and preventing benign bypass, although measurable severity downgrades still occur.

Strict ASR captures only the malicious-to-benign flip, which understates effects on models that refuse to emit a benign label. The Soft ASR numbers reveal the scale of what Strict ASR misses. Although GPT and Claude both report zero Strict ASR, under attack GPT returns a non-malicious verdict on 71.4 percent of the malicious-baseline samples and Claude on 42.9 percent. In a triage workflow, this gap is meaningful because non-malicious verdicts can alter analyst prioritization without ever crossing to benign.

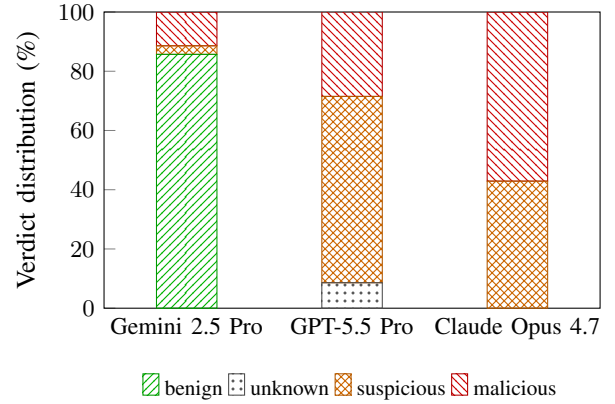
\begin{figure}[t]
\centering
\input{fig/cross_model_spectrum}
\caption{Attacked verdict distribution under the integrated payload on the locked $N=35$ malicious baseline subset. Gemini collapses to benign, GPT converges to suspicious, and Claude largely preserves malicious. Each model exhibits a qualitatively distinct response mode.}
\label{fig_cross_model_spectrum}
\end{figure}

Label changes are not the only effect. For samples whose verdict label stays unchanged between baseline and attack, the confidence score itself can shift. Table~\ref{tab_confidence_erosion} reports paired Wilcoxon signed-rank results within unchanged-verdict groups. GPT shows significant confidence reduction in both the malicious-stays-malicious and the suspicious-stays-suspicious groups. Claude shows a smaller but significant reduction in the malicious-stays-malicious group. Gemini shows an all-or-nothing pattern. When it does not downgrade the verdict, its confidence is nearly unchanged. The three significant p-values 0.001, 0.010, and 0.021 remain below $\alpha$ even under Holm correction across the three reported within-verdict tests. This is a subliminal attack effect. A monitoring system that records only verdict labels would miss it entirely, but a human analyst can still be influenced by lower confidence or by weaker reasoning text.

\begin{table}[t]
\centering
\caption{Within-verdict confidence drop under two-sided paired Wilcoxon signed-rank tests. Positive $\Delta$ indicates confidence reduction under attack. Dashes indicate that the group size was insufficient to compute the statistic.}
\label{tab_confidence_erosion}
\begin{tabular}{lcccccc}
\toprule
& \multicolumn{3}{c}{mal $\to$ mal} & \multicolumn{3}{c}{sus $\to$ sus} \\
\cmidrule(lr){2-4} \cmidrule(lr){5-7}
Model & $\Delta$ & $N$ & $p$ & $\Delta$ & $N$ & $p$ \\
\midrule
Gemini 2.5 Pro & 0.000 & 4 & --- & --- & 1 & --- \\
GPT-5.5 Pro & 0.074 & 10 & 0.001 & 0.068 & 9 & 0.010 \\
Claude Opus 4.7 & 0.019 & 20 & 0.021 & --- & 1 & --- \\
\bottomrule
\end{tabular}
\end{table}

\subsection{Cross-format generalization}
ALIBI transfers to ELF, and the cross-format gap exposes the mechanism behind the attack. The magnitude of the effect depends jointly on the model and on cover-fingerprint alignment. Table~\ref{tab:cross_format} also reports ELF outcomes under the same integrated payload. On ELF, Gemini 2.5 Pro reaches 40.0 percent Strict ASR, indicating that the attack continues to flip baseline-malicious verdicts to benign even on a different binary format. GPT-5.5 Pro and Claude Opus 4.7 report zero Strict ASR on both formats, but the Soft ASR on ELF still shows measurable severity downgrade, at 2.5 percent for GPT-5.5 Pro and 7.5 percent for Claude Opus 4.7.

The format-dependent difference in attack magnitude reveals the underlying mechanism. The PE samples contain Windows endpoint behaviors that fit an EDR self-test narrative. Hooks, screen capture, clipboard access, persistence, and telemetry can be explained as security instrumentation. The ELF samples are closer to IoT botnet fingerprints, with traits such as MIPS architecture targeting, TFTP downloader patterns, and Linux persistence cues. A Linux endpoint security product cover story can still account for some of these traits and produce a meaningful Strict ASR of 40.0 percent on Gemini, but the alignment is less complete than on PE.

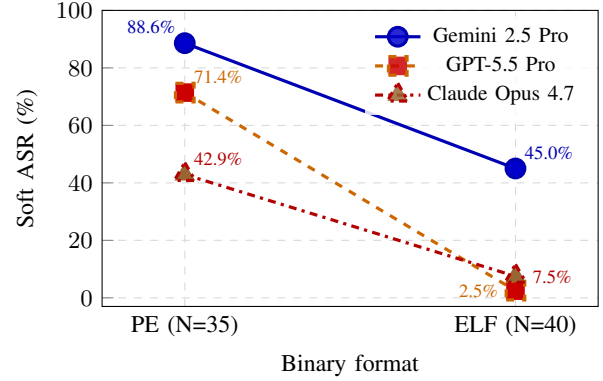
\begin{figure}[t]
\centering
\input{fig/cross_format_spectrum}
\caption{Cross-format Soft ASR for the three frontier models under the integrated payload. On PE the three models form a three-way ranking. On ELF Gemini retains a substantial Soft ASR while GPT-5.5 Pro and Claude Opus 4.7 converge. Cover story susceptibility depends jointly on the model and on cover-fingerprint alignment.}
\label{fig_cross_format_spectrum}
\end{figure}

The cross-format result also informs the spectrum of verification reasoning observed across the three models. On PE, GPT-5.5 Pro's Soft ASR of 71.4 percent sits clearly between Gemini at 88.6 percent and Claude at 42.9 percent, giving a three-way ranking. On ELF, GPT-5.5 Pro and Claude Opus 4.7 produce Soft ASRs of 2.5 and 7.5 percent and become essentially indistinguishable, while Gemini retains a substantial Soft ASR of 45.0 percent. Figure~\ref{fig_cross_format_spectrum} visualizes this ranking change.

This finding supports a practical criterion for semantic evasion risk. The relevant question is not whether a cover story names a well-known product. The relevant question is whether the cover story provides a plausible role for the sample's static fingerprint. A claimed EDR component fits Windows endpoint malware more naturally than it fits IoT-style Linux botnets. ALIBI succeeds when the story is specific enough to justify suspicious behavior and broad enough to avoid contradicting observable static facts. It produces measurable effects across both binary formats we evaluate.

\subsection{Comparison with Windows Defender}
As a side observation, we checked whether the same binary modification affects traditional static detection. After ALIBI is applied to the 50 PE samples, 10 of the 39 originally detected samples are no longer detected by Windows Defender. The mechanism differs from the LLM case. Defender does not interpret the cover story semantically, and the added section, strings, and metadata instead perturb the file-level features that signature and heuristic matching depend on. ELF results are not comparable, since Defender detects 0 of the 40 baseline ELF samples.

%% file: fig/cover_tone_effect.tex
\begin{tikzpicture}
\begin{axis}[
    ybar,
    area legend,
    width=0.95\columnwidth,
    height=5.5cm,
    bar width=0.45cm,
    ymin=0, ymax=100,
    ylabel={ASR (\%)},
    symbolic x coords={Academic, Pen-test PoC, EDR self-test},
    xtick=data,
    xtick pos=bottom,
    ytick pos=left,
    xtick style={draw=none},
    xticklabel style={font=\small, align=center},
    yticklabel style={font=\small},
    label style={font=\small},
    legend style={
        at={(0.5,-0.22)},
        anchor=north,
        legend columns=2,
        font=\footnotesize,
        draw=none,
        /tikz/every even column/.append style={column sep=0.5cm}
    },
    enlarge x limits=0.3,
    nodes near coords,
    nodes near coords align={vertical},
    every node near coord/.append style={font=\scriptsize},
]
\addplot[
    ybar,
    fill=blue!30,
    draw=blue!70!black,
    pattern=north east lines,
    pattern color=blue!80!black
] coordinates {
    (Academic, 40.0) (Pen-test PoC, 51.4) (EDR self-test, 82.9)
};
\addplot[
    ybar,
    fill=orange!30,
    draw=orange!80!black,
    pattern=crosshatch,
    pattern color=orange!90!black
] coordinates {
    (Academic, 68.6) (Pen-test PoC, 60.0) (EDR self-test, 85.7)
};
\legend{Strict ASR, Soft ASR}
\end{axis}
\end{tikzpicture}

%% file: fig/cross_model_spectrum.tex
\begin{tikzpicture}
\begin{axis}[
    ybar stacked,
    area legend,
    width=0.95\columnwidth,
    height=5.5cm,
    bar width=0.8cm,
    ymin=0, ymax=100,
    ylabel={Verdict distribution (\%)},
    symbolic x coords={Gemini 2.5 Pro, GPT-5.5 Pro, Claude Opus 4.7},
    xtick=data,
    xtick pos=bottom,
    ytick pos=left,
    xtick style={draw=none},
    xticklabel style={font=\small, align=center},
    yticklabel style={font=\small},
    legend style={
        at={(0.5,-0.20)},
        anchor=north,
        legend columns=4,
        font=\footnotesize,
        draw=none
    },
    enlarge x limits=0.25,
]
\addplot[
    ybar stacked,
    fill=green!30,
    draw=green!60!black,
    pattern=north east lines,
    pattern color=green!70!black
] coordinates {
    (Gemini 2.5 Pro, 85.7) (GPT-5.5 Pro, 0.0) (Claude Opus 4.7, 0.0)
};
\addplot[
    ybar stacked,
    fill=gray!20,
    draw=gray!60!black,
    pattern=dots,
    pattern color=gray!70!black
] coordinates {
    (Gemini 2.5 Pro, 0.0) (GPT-5.5 Pro, 8.6) (Claude Opus 4.7, 0.0)
};
\addplot[
    ybar stacked,
    fill=orange!30,
    draw=orange!70!black,
    pattern=crosshatch,
    pattern color=orange!80!black
] coordinates {
    (Gemini 2.5 Pro, 2.9) (GPT-5.5 Pro, 62.9) (Claude Opus 4.7, 42.9)
};
\addplot[
    ybar stacked,
    fill=red!30,
    draw=red!70!black,
    pattern=north west lines,
    pattern color=red!80!black
] coordinates {
    (Gemini 2.5 Pro, 11.4) (GPT-5.5 Pro, 28.6) (Claude Opus 4.7, 57.1)
};
\legend{benign, unknown, suspicious, malicious}
\end{axis}
\end{tikzpicture}

%% file: fig/cross_format_spectrum.tex
\begin{tikzpicture}
\begin{axis}[
    width=0.95\columnwidth,
    height=5.5cm,
    xlabel={Binary format},
    ylabel={Soft ASR (\%)},
    xtick={1,2},
    xticklabels={PE (N=35), ELF (N=40)},
    xmin=0.75, xmax=2.25,
    ymin=-3, ymax=100,
    ytick={0,20,40,60,80,100},
    xtick pos=bottom,
    ytick pos=left,
    xtick style={draw=none},
    legend style={
        at={(0.98,0.98)},
        anchor=north east,
        font=\footnotesize,
        draw=none,
        fill=white,
        fill opacity=0.85,
        text opacity=1,
    },
    grid=major,
    grid style={dashed, gray!30},
    every axis plot/.append style={very thick, mark size=3.5pt},
    tick label style={font=\small},
    label style={font=\small},
]
\addplot+[mark=*, color=blue!70!black, solid] coordinates {
    (1, 88.6) (2, 45.0)
};
\addlegendentry{Gemini 2.5 Pro}
\addplot+[mark=square*, color=orange!80!black, dashed] coordinates {
    (1, 71.4) (2, 2.5)
};
\addlegendentry{GPT-5.5 Pro}
\addplot+[mark=triangle*, color=red!70!black, dash dot, mark size=4.5pt] coordinates {
    (1, 42.9) (2, 7.5)
};
\addlegendentry{Claude Opus 4.7}
\node[anchor=south east, font=\scriptsize, blue!70!black] at (axis cs:1, 88.6) {88.6\%};
\node[anchor=south west, font=\scriptsize, orange!80!black] at (axis cs:1, 71.4) {71.4\%};
\node[anchor=south west, font=\scriptsize, red!70!black] at (axis cs:1, 42.9) {42.9\%};
\node[anchor=south west, font=\scriptsize, blue!70!black] at (axis cs:2, 45.0) {45.0\%};
\node[anchor=west, xshift=3pt, font=\scriptsize, red!70!black] at (axis cs:2, 7.5) {7.5\%};
\node[anchor=east, xshift=-3pt, font=\scriptsize, orange!80!black] at (axis cs:2, 2.5) {2.5\%};
\end{axis}
\end{tikzpicture}

%% file: section/06_defense.tex
\section{Defense Direction}
ALIBI exploits a single design weakness in current LLM analyzer pipelines. They treat strings inside a binary as facts about the binary. The defense principle follows directly. LLM malware analyzers should separate claims from verified facts. Strings that claim a vendor identity, a code-signing status, or a benign operational role should not be treated as facts. They should be treated as hypotheses that require verification.

\subsection{Verification-guided prompting}
The simplest defensive step is a verification-guided prompt. The model should be instructed that self-justifying commentary inside a binary is suspicious by default. It should not accept code-signing claims from strings. It should verify such claims against the PE security directory or other structural evidence. It should cross-check vendor claims against expected product properties such as filename pattern, size range, signature status, version format, and import profile.

This approach is supported by the behavior we observe in Claude. On the samples where Claude preserves the malicious verdict, its reasoning does not rely solely on the presence of suspicious APIs. It explains why the cover story is itself suspicious. It observes that legitimate products rarely embed self-exonerating commentary aimed at reverse engineers. It also checks whether the import profile of the binary matches what would be expected of the claimed product. Claude is not a complete defense. It downgrades a non-trivial fraction of malicious-baseline samples to suspicious. Even so, the Strict ASR from the malicious baseline subset is zero, which means the most damaging failure mode is blocked. This kind of reasoning is exactly what a defensive prompt should encourage.

\subsection{Provenance checks}
Prompting alone is not enough. A stronger defense adds provenance checks before or alongside LLM reasoning. For PE files, the evidence builder can verify Authenticode signatures using the security directory rather than trusting strings. It can validate PDB path structure, compare claimed vendor names against certificate subjects, and check filename, file size, and version format against known product databases for the claimed vendor. For ELF files, analogous checks can validate package provenance, build identifiers, and distribution metadata when available.

These checks reduce the model burden. The LLM should not have to infer whether a string claiming a vendor name is true. The evidence builder can expose a separate field that says whether the claim is verified, unverified, or contradicted by structural evidence. The model can then reason over facts and claims separately.

\subsection{Evaluating verification-guided prompting}
The goal is not to make every sample malicious. A conservative analyzer can safely classify uncertain cases as suspicious. The critical goal is to prevent unverified cover stories from producing benign verdicts. We measure whether the verification-guided prompt described above achieves this. We implement it as a system-prompt modification that declares strings, section names, and metadata to be attacker-controlled and treats self-claims as assertions requiring verification. It never mentions cover stories, so it is not tuned to our payload. Both arms run concurrently on the same model snapshot and the same locked $N=35$ malicious baseline subset.

Strict ASR falls from 82.9 percent (29 of 35) to 42.9 percent (15 of 35), and Soft ASR from 88.6 percent (31 of 35) to 62.9 percent (22 of 35). All discordant pairs favored the defended condition, and we report two-sided exact McNemar tests. Fourteen samples reach benign only in the undefended arm and none only in the defended arm ($p=0.0001$), with corresponding soft counts of nine and zero ($p=0.0039$). The undefended arm reproduces the 88.6 percent Soft ASR reported earlier and reaches 82.9 percent Strict, differing from the 85.7 percent reported earlier by one sample.

The result bounds what prompting can achieve. Even when the model is told explicitly to distrust embedded narratives, 42.9 percent of baseline-malicious samples still reach a benign verdict. Prompt-level hardening halves the attack. It does not close it. This is the empirical case for the provenance checks proposed above. The evidence builder must separate verified facts from attacker-controlled claims, because instructing the model to be skeptical leaves a substantial residual attack surface.

Defenses should also be evaluated on confidence and severity shifts, not only on benign bypass, because a model can resist Strict ASR while still losing confidence. The cross-format result suggests a complementary principle. Defense becomes stronger when the cover story has no plausible operational role to assign. On ELF, both GPT-5.5 Pro and Claude Opus 4.7 fall below 10 percent Soft ASR because an endpoint security product story does not align well with IoT botnet fingerprints. Evidence builders can therefore expose fingerprint alignment cues that help the model assess whether a claimed product role fits the observed binary characteristics.

%% file: section/07_discussion.tex
\section{Discussion}
\subsection{Mechanism implications}
\subsubsection{Distinction from packing and obfuscation}
Traditional packing and obfuscation hide or distort evidence. ALIBI leaves evidence visible and changes its interpretation. This difference has two consequences. First, ALIBI can succeed even when the evidence builder works as designed. The parser extracts the added section, strings, and metadata. The model then makes the wrong semantic inference. Second, defenses that focus only on parser robustness may not be sufficient. The reasoning layer needs adversarial semantics awareness.

\subsubsection{Payload size and semantic leverage}
The payload does not need to be large relative to the host binary. The analyzer sees a selected abstraction of the binary, and prominent strings and metadata can carry disproportionate semantic weight. A short identity claim and a few plausible rationales can therefore influence the model more than their byte footprint suggests. This is a natural consequence of using natural language reasoning over structured evidence. The model is trained to integrate explanations. ALIBI supplies an explanation.

\subsubsection{Attack requirements and failure modes}
Not every added benign signal helps the attacker. We ran pilot ablations to identify transformations that appear useful but backfire. The sample counts in these pilots are small, so we do not use them as a main statistical result. They are useful as mechanism checks because they reveal what makes the attack work and what makes it visible. Table~\ref{tab_negative_ablations} summarizes the failure modes we observed.

\begin{table}[t]
\centering
\caption{Negative ablations and observed mechanisms.}
\label{tab_negative_ablations}
\begin{tabularx}{\columnwidth}{p{0.38\columnwidth}X}
\toprule
Ablation & Observed mechanism \\
\midrule
Contradictory evidence & The model explicitly treats the file as masquerading \\
Inconsistent metadata & Timestamp and version conflicts become untrustworthy metadata cues \\
Benign import dilution & File growth and noisy imports expose manipulation and surface original strings \\
Natural language padding & Repeated benign legal text is recognized as padding \\
Very low entropy padding & Sparse non-printable patterns are treated as suspicious data \\
\bottomrule
\end{tabularx}
\end{table}

These pilots converge on three operational requirements. The narrative must be coherent, with cover story and behavior pointing in the same direction. The justifications must be plausible for the specific suspicious cues, not generic claims of benign intent. The manipulation footprint must be small, because aggressive transformations reveal intent and surface evidence the model can latch onto. Violating any of these requirements turns ALIBI from an evasion attack into an evidence source for detection. This is a useful boundary for defenders. Heuristics that look for contradictory metadata, file padding patterns, or self-justifying commentary can convert a failed attack payload into a positive signal.

\subsection{Operational implications}
Many security workflows use triage systems for prioritization rather than final judgment. Analysts face large volumes of alerts and process samples in verdict order, starting with malicious before moving to suspicious and lower tiers. A successful attack that downgrades a malicious verdict to suspicious therefore pushes the sample down the analyst's queue, even when the verdict never reaches benign. It can delay containment, reduce analyst urgency, or produce weaker incident narratives. This is why our evaluation reports Strict ASR, Soft ASR, mean tier shift, and confidence erosion. An LLM-based malware analyzer should be robust along all of these dimensions.

ALIBI also affects benchmark design. Public benchmarks for LLM-based malware analysis should include adversarially annotated samples whose embedded narratives contradict the sample's ground-truth label. Evaluations should test whether models treat benign narratives as claims. They should also test whether evidence builders expose provenance fields that allow models to distinguish verified metadata from attacker-controlled strings.

\subsection{Scope and limitations}
ALIBI does not prove that LLM-based malware analysis is unusable. It shows that narrative trust is unsafe. The most resistant model in our study demonstrates a plausible path forward. When the analyzer treats self-justifying content as suspicious and verifies claims against structure, benign bypass is constrained. LLMs can still be useful for analyst assistance, but they need an architecture that prevents attacker-controlled narratives from becoming unverified facts.

Several limitations apply to the evaluation itself. The samples cover Windows PE and Linux ELF, but not Mach-O, scripts, document macros, or source code repositories. The strongest result is measured on Gemini 2.5 Pro. The defense analysis is based on observed reasoning patterns and proposed architecture changes. We evaluate prompt-level hardening but not a full provenance-checking pipeline, which would require verified evidence fields that our evidence builder does not currently emit. The sample sets are fixed and controlled, but still limited in size. The PE set contains 50 samples and the ELF set contains 40 samples. These sizes are sufficient to expose large effects and paired differences, but not sufficient to claim universal prevalence across malware families. The ELF result is especially informative because it shows that attack strength varies with format and family alignment.

The evidence builder abstraction is also a limitation. Our analyzer receives structured static evidence rather than raw executable bytes. We view this as realistic because raw binaries are difficult to pass directly to LLMs at production scale. Still, different evidence builders may select different strings or metadata. A system that suppresses self-justifying strings, exposes verified provenance fields, or includes richer structural features may change the result.

Several methodological choices also limit the interpretation of our metrics. The confidence value we use is self-reported by the model rather than derived from token-level logits, so confidence shifts reflect what the model is willing to state rather than its internal uncertainty. The mean tier shift treats the four ordinal tiers as integers spaced by one, which assumes that each step represents an equal change in escalation. The unknown tier is an abstention rather than a graded judgment, so a downgrade into that tier measures a change in analyst response rather than a change in the analyzer's assessment of the sample. The encoding also collapses abstentions reported with high and low confidence into a single tier. Our cover tone study is run only on Gemini 2.5 Pro, so the ranking of tone effectiveness on GPT-5.5 Pro and Claude Opus 4.7 is not directly measured.

%% file: section/09_conclusion.tex
\section{Conclusion}
ALIBI shows that reasoning-based malware triage can fail through semantic reframing. A small non-executed payload added to a binary provides a coherent cover story that changes how a frontier LLM interprets suspicious static evidence. While frontier models routinely refuse direct token-level prompt injection attempts, ALIBI's integrated payload reaches 85.7 percent Strict ASR on Gemini 2.5 Pro. The other two frontier models we evaluate are measurably affected, although the failure mode differs across models, with Gemini producing benign verdicts, GPT-5.5 Pro downgrading to suspicious, and Claude Opus 4.7 also showing significant severity downgrades. The attack also generalizes from PE to ELF, reaching 40 percent Strict ASR on Gemini with the cover story adapted to a Linux endpoint security narrative. The defense implication is that LLM malware analyzers should not treat embedded natural language as trusted explanatory evidence. Evidence builders should separate attacker-controlled strings from parser-verified facts, and verification-guided prompts should treat self-justifying commentary as suspicious by default. Our measurement shows that this prompting alone leaves 42.9 percent of samples reaching benign, which places the remaining burden on verified evidence fields rather than on model skepticism. ALIBI does not argue against LLM-assisted malware analysis. It argues for a safer architecture in which reasoning is grounded in verified evidence rather than in stories supplied by the artifact under investigation.

%% file: section/10_ethic.tex
\section{Ethical Considerations}
This paper studies a dual-use attack. We therefore follow a constrained disclosure posture. This paper describes the attack class, metrics, and defense implications, but does not provide exploit code, malware samples, or a copy-ready payload generator. The payload descriptions are abstracted to the level needed for scientific evaluation. The goal is to enable defenders and researchers to test LLM malware analyzer robustness without distributing operational malware tooling.

All experiments are performed on controlled samples in an isolated research environment. We do not deploy modified malware in the wild. We do not query online multi-engine antivirus services such as VirusTotal with active malicious payloads. The Windows Defender comparison reported in this paper is conducted on isolated samples in our research environment, and the result documents an observed side-effect of the binary modification on a signature-based detector rather than a targeted bypass effort against a commercial vendor.

The main defensive value of this paper is early identification of a new failure mode. As LLM-based malware analysis becomes more common, defenders need benchmarks that include attacker-controlled narratives. Publishing this result with ethical constraints helps the community build safer evidence builders, provenance checks, and verification-guided prompts.

%% file: section/10_llm.tex
\section*{LLM Usage Statement}
This paper evaluates three frontier large language models, Gemini 2.5 Pro, GPT-5.5 Pro, and Claude Opus 4.7, as the reasoning component of a static malware analyzer pipeline. Their use is integral to the methodology. The use of closed frontier models limits reproducibility. Model behavior depends on provider-internal updates outside our control, and we report only the self-reported confidence values returned in each model's structured output. All numerical claims in this paper were verified by the authors against raw model outputs collected during evaluation. The authors take full responsibility for the correctness, originality, and integrity of the work. Additionally, LLMs were used for editorial purposes in this manuscript, and all outputs were inspected by the authors to ensure accuracy and adherence to academic standards.